\documentclass[aps,prl,twocolumn,showpacs,superscriptaddress,floatfix]{revtex4-1}  % for review and submission
\usepackage[english]{babel}
\usepackage{graphicx}
\usepackage{dcolumn}
\usepackage{amsmath}
\usepackage{natbib}
\usepackage{bm}
\usepackage[utf8]{inputenc}

\newcommand{\mum}{\mu\mbox{m}}
\newcommand{\sign}{\mathop{\rm sign}}
\renewcommand{\vec}[1]{\mathbf{#1}}

\begin{document}

\title{Landau level spectroscopy of electron-electron interactions in graphene  }
%{Coulomb interactions effects in graphene
%as observed by magneto-Raman scattering}
%as seen by inter-Landau-level excitations}

\author{C.~Faugeras}
\affiliation{Laboratoire National des Champs Magn\'etiques
Intenses, CNRS, (UJF, UPS, INSA), BP 166, 38042 Grenoble, Cedex 9,
France}

\author{S.~Berciaud}
\affiliation{Institut de Physique et Chimie des Mat\'eriaux de
Strasbourg and NIE, UMR 7504, Universit\'e de Strasbourg and CNRS,
BP43, 67034 Strasbourg Cedex 2, France}

\author{P.~Leszczynski}
\affiliation{Laboratoire National des Champs Magn\'etiques
Intenses, CNRS, (UJF, UPS, INSA), BP 166, 38042 Grenoble, Cedex 9,
France}
\author{Y.~Henni} \affiliation{Laboratoire National des Champs Magn\'etiques
Intenses, CNRS, (UJF, UPS, INSA), BP 166, 38042 Grenoble, Cedex 9,
France}
\author{K.~Nogajewski} \affiliation{Laboratoire National des Champs Magn\'etiques
Intenses, CNRS, (UJF, UPS, INSA), BP 166, 38042 Grenoble, Cedex 9,
France}
\author{M.~Orlita} \affiliation{Laboratoire National des Champs Magn\'etiques
Intenses, CNRS, (UJF, UPS, INSA), BP 166, 38042 Grenoble, Cedex 9,
France}

\author{T.~Taniguchi} \affiliation{National Institute for Material Science, 1-1 Namiki, Tsukuba, Japan}
\author{K.~Watanabe} \affiliation{National Institute for Material Science, 1-1 Namiki, Tsukuba, Japan}

\author{C.~Forsythe} \affiliation{Department of Physics, Columbia University, New York, NY 10027, USA}
\author{P.~Kim} \affiliation{Department of Physics, Columbia University, New York, NY 10027, USA}

\author{R.~Jalil} \affiliation{School of Physics and Astronomy, University of
Manchester, Manchester, M13 9PL, United Kingdom}
\author{A.K.~Geim} \affiliation{School of Physics and Astronomy, University of
Manchester, Manchester, M13 9PL, United Kingdom}

\author{D.M.~Basko}\email{denis.basko@lpmmc.cnrs.fr}
\affiliation{Universit\'e Grenoble 1/CNRS, Laboratoire de Physique
et de Mod\'elisation des Milieux Condens\'es (UMR 5493), B.P. 166,
38042 Grenoble, Cedex 9, France}

\author{M.~Potemski}\email{marek.potemski@lncmi.cnrs.fr}
\affiliation{Laboratoire National des Champs Magn\'etiques
Intenses, CNRS, (UJF, UPS, INSA), BP 166, 38042 Grenoble, Cedex 9,
France}

\date{\today}

\begin{abstract}
We present magneto-Raman scattering studies of electronic inter
Landau level excitations in quasi-neutral graphene samples with
different strengths of Coulomb interaction. The band velocity
associated with these excitations is found to depend on the
dielectric environment, on the index of Landau level involved, and
to vary as a function of the magnetic field. This contradicts the
single-particle picture of non-interacting massless Dirac
electrons, but is accounted for by theory when the effect of
electron-electron interaction is taken into account. Raman active,
zero-momentum inter Landau level excitations in graphene are
sensitive to electron-electron interactions due to the
non-applicability of the Kohn theorem in this system, with a
clearly non-parabolic dispersion relation.
\end{abstract}

 \maketitle

Single-particle electronic states in graphene ostentatiously
follow the dispersion of massless fermions, described by the Weyl
equation in which the speed of light~$c$ is simply scaled down to
the band velocity $v \sim c/300$. Recently, however, more and more
attention is paid to the effects of interactions and, in
particular, to a modification of the dispersion relations and
excitation spectra of quasi-particles induced by electron-electron
interactions~\cite{Elias2011, Chen2014, Basov2014}. Indeed,
graphene, and in particular pristine graphene, can hardly be
considered as a weakly interacting
system~\cite{Gonzalez1993,Hofmann2014,Kotov2012}. The
dimensionless interaction strength (the ratio between typical
Coulomb and kinetic energies), which is rather small in genuine
systems of quantum electrodynamics, $\alpha\approx{1}/137$ (the
fine-structure constant), appears to be sizable in graphene,
$(c/v)\alpha\approx{2}$. Screening (by a dielectric and/or
conducting environment) naturally alters the strength of the
electron-electron interaction in graphene, depending on its actual
surrounding (substrate) and/or on the degree of departure from
charge neutrality (electron/hole concentration). In a uniform
dielectric environment characterized by a dielectric
constant~$\varepsilon$, the effective fine-structure constant is
$\alpha_\varepsilon=(c/v)(\alpha/\varepsilon)$. The
renormalization of graphene bands by electron-electron
interactions has been mostly studied in the absence of magnetic
fields~\cite{Basov2014}, whereas the anticipated
effects~\cite{Iyengar2007,Bychkov2008,Lozovik2012,Shizuya2010,Orlita2010}
of these interactions in the regime of Landau quantized energy
levels have been little explored so far~\cite{Jiang2007,Chen2014}.

In the present work, we investigate the effects of
electron-electron interactions in graphene subjected to quantizing
magnetic fields by probing its inter-Landau-level excitations with
magneto-Raman scattering
experiments~\cite{Faugeras2011,Berciaud2014}. We have studied
three graphene systems with different dielectric environments. The
non-interacting Dirac-like description of electronic states fails
to account for the full set of our experimental observations. The
velocity parameter, which we associate with each LL transition, is
not a single value, but: (i) changes with the effective dielectric
constant expected in our samples; the departure from the
non-interacting picture is most pronounced for suspended graphene,
weaker for graphene encapsulated in hexagonal boron nitride and
rather small for graphene on graphite, (ii) varies logarithmically
with the magnetic field, (iii) is higher for transitions involving
higher LLs. These observations can be qualitatively described in
the Hartree-Fock
approximation~\cite{Iyengar2007,Bychkov2008,Lozovik2012} or by the
first-order perturbation theory (FOPT) in
$\alpha_\varepsilon$~\cite{Gonzalez1993, Shizuya2010}. In
particular, these calculations yield no full cancellation between
vertex and self-energy corrections, implying violation of the Kohn
theorem~\cite{Kohn1961} for the Dirac spectrum. Notably, the
vertex corrections invert the tendency of lowering the electron
velocity with energy, resulting from the self-energy terms, which
accounts for feature~(iii); see also ~\cite{Jiang2007,Chen2014}.
However, FOPT fails on the quantitative level when
$\alpha_\varepsilon$ is not small. Beyond FOPT, the leading terms
in $\ln(B)$ can be addressed by the random-phase approximation
(RPA) ~\cite{Gonzalez1999} (see also ~\cite{Hofmann2014}), which
turns out to match the experimental results quite well . Under
some additional assumptions, we estimate two relevant parameters,
the band width and bare band velocity, which define the
renormalized electronic dispersion.

Conventional absorption spectroscopy of inter-LL transitions in
graphene~\cite{Orlita2010} is restricted to far-infrared spectral
range ($\lambda \sim 100\:\mum$) and does not offer the necessary
spatial resolution, otherwise required for probing small graphene
flakes. Better resolution is offered by visible light techniques,
such as Raman scattering which is our method of choice. The
possibility of observing Raman scattering from purely electronic,
inter LL excitations~\cite{Kashuba2009,Faugeras2011,Berciaud2014}
is a recent addition to the wide use of Raman scattering spectra
of phonons for the characterization of different graphene
structures~\cite{Ferrari2013,Malard2009}. We studied three
distinct graphene systems: suspended graphene (G-S), graphene
encapsulated in hexagonal boron nitride (G-BN) and graphene on
graphite (G-Gr). G-S was suspended over a circular pit ($8\:\mum$
in diameter) patterned on the surface of an Si/SiO$_2$ substrate
(see Ref.~\cite{Berciaud2009, Berciaud2014} for details of sample
preparation). The G-BN structure consists of a graphene flake
transferred onto a $\sim 50$~nm thick layer of hBN and then
covered by another hBN flake of the same thickness, all together
placed on an Si/SiO$_2$ substrate (see Ref.~\cite{Dean2010} for
details on a similar structure). The G-Gr flake was identified on
the surface of freshly exfoliated natural graphite via mapping the
Raman scattering response at a fixed magnetic field and searching
for the position with the spectral features characteristic for
graphene(see Ref.~\cite{Faugeras2014} for details of the
procedure). The experimental arrangements (see also
Ref.~\cite{Faugeras2011,Berciaud2014}) permitted Raman scattering
experiments in magnetic fields up to $14$~T (supplied by a
superconducting coil, data collected for G-BN) or up to $29/30$~T
(supplied by a resistive magnet, data collected for G-S and G-Gr),
at low temperatures ($4$~K) and with a spatial resolution of $\sim
1\:\mum$ (diameter of the laser spot on the sample).

Magneto-Raman scattering spectra on G-Gr and G-S have been
measured using the 514.5 nm line of an Ar$^{+}$ laser for the
excitation, in a simple, unpolarized light configuration.
Experiments on G-BN were more demanding, due to a superfluous
scattering/emission background originating from the hBN layers. To
better resolve the electronic response from the G-BN species,
laser excitation of a longer wavelength ($\sim780$~nm) was chosen
and the polarization resolved technique was implemented in the
configuration of the circularly-polarized excitation beam and the
back-scatted Raman signal, both of the same
helicity~\cite{Kuhne2012,Kashuba2009}. All experiments were
performed using a laser power of $\sim1$~mW at the sample. The
sample was mounted on an X-Y-Z micro-positioning stage, which
enabled us to map the Raman scattering response over the sample
surface and to locate the graphene flake. The performance of our
set-up is limited to the detection of Raman scattering signals
exceeding an energy of $\sim 400$~cm$^{-1}$ from the laser line,
due to various, spectral blocking elements/filters incorporated in
the system. The magneto-Raman spectra have been recorded one after
another while slowly sweeping the magnetic field. Typically, each
spectrum was accumulated over a time interval during which the
magnetic field was changed by $\sim 0.1$~T. Though the adequate
electrical characterization of the investigated samples was not
possible, we assume here that all our three graphene flakes are
not far from being neutral systems; this is supported by many
other studies of similar
structures~\cite{Bolotin2008,Berciaud2009,Dean2010,Neugebauer2009}.

\begin{figure}
\includegraphics[width=0.43\textwidth]{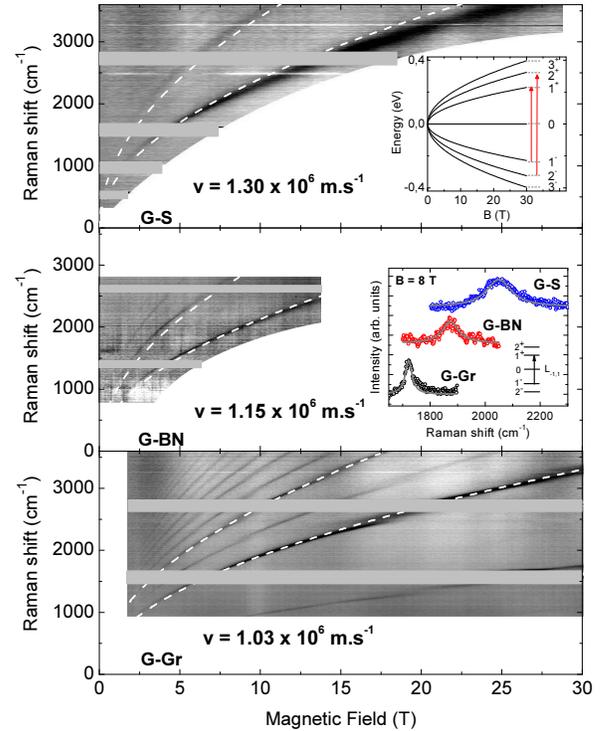}
\caption{\label{fig1} Gray scale map of the electronic response in
the magneto-Raman scattering spectra response as a function of the
magnetic field ($B$) for suspended graphene (G-S, top panel),
graphene encapsulated in hBN (G-BN, middle panel) and graphene on
graphite (G-Gr, bottom panel). Each map represents the collection
of differential spectra (the $B=0$ spectrum has been subtracted
from each spectrum measured in a magnetic field $B$). Dashed white
lines account for the $B$-evolutions of two most pronounced inter
LL excitations, expected within a single particle approach, when
assuming a $B$-independent, but different for each sample, band
velocity $v$. Upper inset: scheme of the Landau level fanchart in
graphene; two, L$_{-1,1}$ and L$_{-2,2}$ inter Landau level
excitations, well visible in the experiments on all samples, are
indicated by red arrows. Middle panel inset: characteristic
spectra due to $L_{-1,1}$ excitations measured at $B=8$~T for G-Gr
(black symbols), G-BN (red symbols) and G-S (blue symbols). Gray
curves are Lorentzian functions. Whitish horizontal ribbons on
gray scale maps are to mask the (residual) contribution to Raman
scattering due to phonons  of graphene/graphite at $\sim1600$ and
$\sim2700$cm$^{-1}$ as well as of Si substrate ( $\sim520$ and
$\sim1000$cm$^{-1}$ at of hBN at $\sim1370$cm$^{-1}$.}
\end{figure}

Besides the well-known spectral peaks due to
phonons~\cite{Ferrari2006,Ferrari2013,Malard2009}, the
magneto-Raman scattering spectra of each of our graphene samples
shows other, well-resolved peaks due to electronic inter-LL
excitations, whose energies depend distinctly on the magnetic
field. Those features are central for the present work. A
collection of the related experimental data is presented in
Fig.~\ref{fig1}. In the zero-order approximation (single particle
approach), the electronic dispersion in graphene is conical,
$E(k)=\pm{v}k$. When a magnetic field~$B$ is applied
perpendicularly to the graphene plane, the continuous energy
spectrum transforms into a series of discrete Landau levels
($L_{\pm n}$) with energies $E_{\pm n} = \pm \sqrt{2n}\,\hbar v /
l_B$ (here $n=0,1,2,\ldots$, and $l_B=\sqrt{\hbar/(eB)}$~is the
magnetic length); see upper inset to Fig.~\ref{fig1}. We limit our
considerations to the so-called symmetric $L_{-n} \rightarrow
L_{n}$ ($L_{-n,n}$) inter LL transitions, which are expected to
dominate the electronic Raman scattering response of graphene
\cite{Kashuba2009}, and appear at energies
$\hbar\omega_{-n,n}=2E_n$, approximately. Tracing the
$\hbar\omega_{-n,n}(B)$ dependences on top of gray-scale maps
presented in Fig.~\ref{fig1}, we recognize two transitions in the
spectra of both G-S and G-BN, $L_{-1,1}$ and $L_{-2,2}$. G-Gr
shows much richer spectra: a larger number of symmetric $L_{-n,n}$
transitions (at least up to $n=5$) as well as other, asymmetric,
$\Delta |n|=1$ transitions. These latter transitions were
predicted to be weakly allowed~\cite{Kashuba2009} but are
nevertheless well seen in G-Gr. We believe that the electronic
quality (mobility) is the crucial factor which influences the
richness of electronic Raman spectra of graphene. Aiming at a
systematic study of the spectra in different dielectric
environments, we focus on the $L_{-1,1}$ and $L_{-2,2}$ which are
clearly seen in all three cases. In all our samples, the
$L_{-1,1}$ transition starts to be visible at magnetic fields as
low as $\sim2.5$~T, at energies $E_{ons} \sim 1000$~cm$^{-1}$.
This observation defines the upper bound for the Fermi energy,
$E_{F}< E_{ons}/2 \simeq 60$~meV, and confirms a relatively low
doping in the studied graphene structures.

From the inspection of the $\hbar \omega_{-n,n} (B)$ traces drawn
in Fig.~\ref{fig1}, we easily identify the measured transitions,
but at the same time notice some inconsistencies. First of all,
using such an approximate data modelling we are forced to use
different velocities for each of our graphene samples: $v$ is set
to $1.30\times 10^6$~m/s, $1.15\times 10^6$~m/s, and $1.03\times
10^6$~m/s for G-S, G-BN, and G-Gr, respectively. The effect of
different mean velocities for each graphene specimen is directly
visualized in the lower inset to Fig.~\ref{fig1}: at fixed $B$ but
for different samples the $L_{-1,1}$ transitions appear at clearly
distinct energies. Moreover, the above $v$ parameters can only be
considered as the mean velocity values, averaged over different
transitions and over the range of magnetic fields applied: note
e.~g. rather pronounced deviations between the white traces and
the central peak positions for G-S (top panel of Fig.~\ref{fig1}).

\begin{figure}
\includegraphics[width=0.39\textwidth]{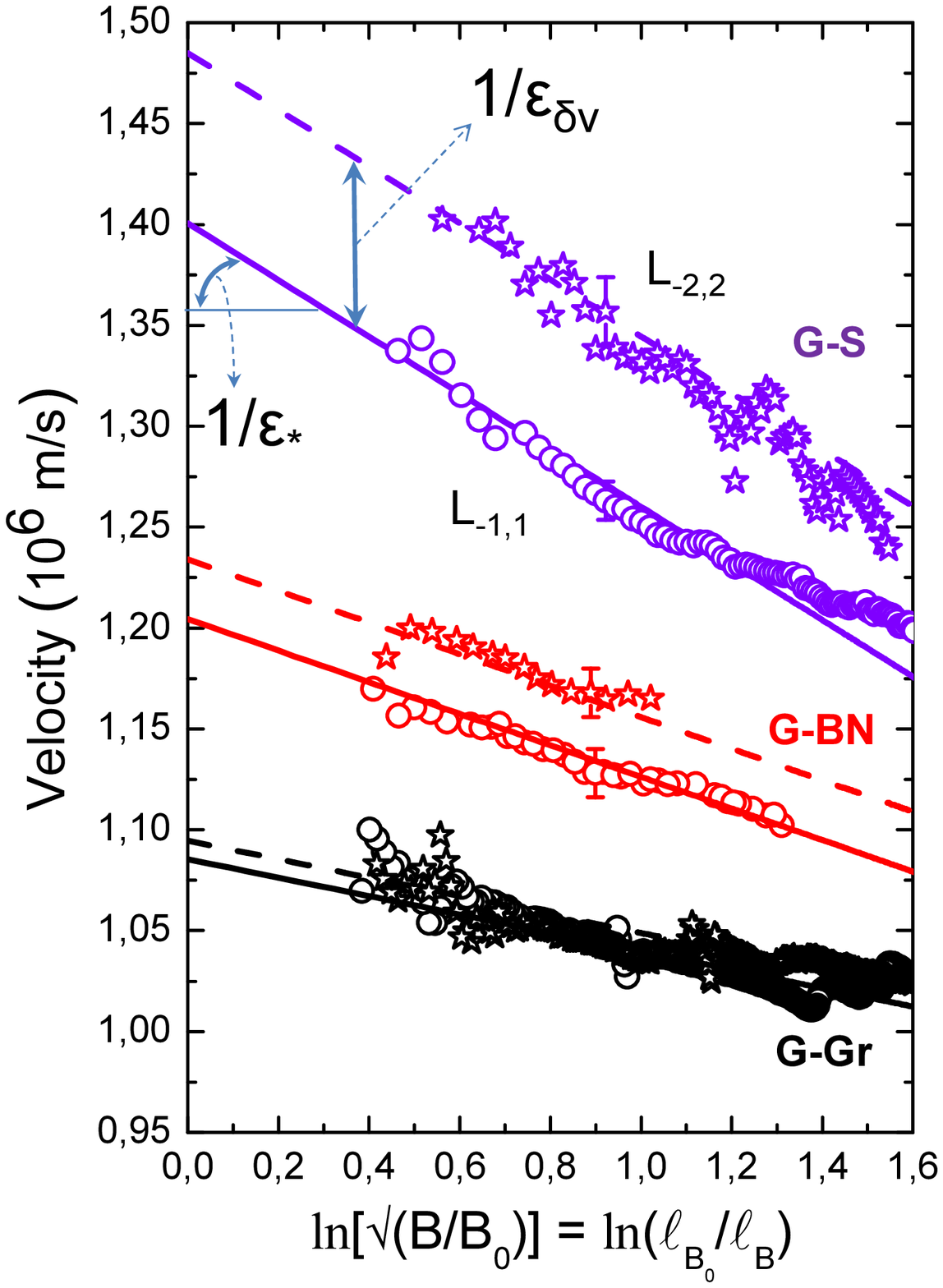}
\caption{\label{fig2} Magnetic field dependence ($B _{0}=$~1~T) of
the velocities associated with L$_{-1,1}$ and L$_{-2,2}$ inter
Landau level excitations shown with, correspondingly, open circles
(solid lines) and open stars (dashed lines), as derived from the
experiment (data modelling), for G-S, G-hBN and G-Gr specimens.
Straight lines follow the Eq.~(\ref{vnPHE=}), (see also
Eq.~(\ref{vnFOPT=}), with $\varepsilon_{*}= 3.9$, $7.0$, and $12$
for G-S, G-hBN and G-Gr species, and the corresponding values for
$\varepsilon_{\delta v} = 1.3$, $3.7$ and $12$}.
\end{figure}

The shortcomings of the above data modelling are emphasized in
Fig.~\ref{fig2}, the central figure of this paper. Data points
(symbols) in this figure represent the velocity parameter which we
associate with each observed transition and at each value of the
magnetic field applied: $v_n^{exp}=
\omega_{-n,n}^{exp}l_B/\sqrt{8n}$, where
$\hbar\omega_{-n,n}^{exp}$ are the measured transition energies
(central positions of the Raman scattering peaks). In the
non-interacting case, all these velocities should collapse onto
one single value. The extracted velocities $v_{n}^{exp}$, for
different $L_{-n,n}$ transitions and for our three graphene
structures, are plotted in Fig.~\ref{fig2} as functions of
$\ln(l_{B_0}/l_B)=\ln\sqrt{B/B_0}$, where the reference value of
the magnetic field has been arbitrarily set to $B_0=1$~T. Each set
of $v_{n}^{exp}$ versus $\ln(l_{B_0}/l_B)$ data can be fairly
approximated by a linear function. The $v_n^{exp}$-traces are
parallel within a given graphene structure but show different
slopes for different samples. These features point towards the
effects of renormalization of the electronic velocity and of
energies of inter-LL transitions by the electron-electron
interactions.

For neutral graphene at $B=0$, the FOPT in
$\alpha_\varepsilon=(c/v_{0})(\alpha/\varepsilon)$ gives the
correction to the velocity~$v$~\cite{Gonzalez1993},

\begin{equation}\label{vFOPT=}
\frac{v}{v_{0}}=1-\frac{\alpha_\varepsilon}{4}\,\ln\frac{|E|}{W},
\hspace{0.8cm} v=v_{0}-\frac{\alpha c}{4
\varepsilon}\,\ln\frac{|E|}{W} ,
\end{equation}

which depends on the electron energy $E$, counted from the Dirac
point. Here $v_{0}$ is the bare velocity, and $W$ is the
high-energy cutoff which is of the order of the electronic
bandwidth (a few eV). The dielectric constant~$\varepsilon$ can be
taken as that of the surrounding medium for the suspended and
encapsulated graphene In a finite magnetic field, the FOPT
calculation of the correction to the transition energy
$\hbar\omega_{-n,n}$, performed analogously to
Ref.~\cite{Shizuya2010}, gives the following correction to the
velocity (see Supplementary Information for details):

\begin{equation}\label{vnFOPT=}
v_n\equiv \frac{\omega_{-n,n}l_B}{\sqrt{8n}} =  v_{0} +
\frac{\alpha c}{4\varepsilon} ( \mathcal{L} - \ln
\frac{l_{B_0}}{l_B}) + \frac{\alpha c}{4\varepsilon} C_n
\hspace{0.5cm}
\end{equation}

where $\mathcal{L}=\ln(Wl_{B_0}/\hbar{v}_0)$ is a constant
resulting from our choice of $l_{B_{0}}$ to set the horizontal
scale in Fig.~\ref{fig2}, and the numerical coefficients are
$C_1=-0.398$ and $C_2=-0.197$. Note that the coefficient in front
of the logarithm (which determines the slopes in Fig.~\ref{fig2})
is the same in Eqs.~(\ref{vFOPT=}) and~(\ref{vnFOPT=}). This is
due to the fact that the leading logarithmic term in
Eq.~(\ref{vnFOPT=}) can be obtained by simply replacing~$E$ in
Eq.~(\ref{vFOPT=}) by the bare LL energy
$E_n^0=\sqrt{2n}\,\hbar{v}_0/l_B$. On the other hand, the
coefficients~$C_n$ include both self-energy and vertex
corrections, and have to be calculated explicitly.

Eq.~(\ref{vnFOPT=}) accounts qualitatively for the main
experimental trends seen in Fig.~\ref{fig2}. For each sample
($\varepsilon$), the dependences of $v_n^{exp}$ versus
$\ln(l_{B_0}/l_B)$ represent a set of parallel lines. The slope of
these lines ($\propto 1/\varepsilon$ according to
Eq.~(\ref{vnFOPT=})) correlates with the expected, progressive
increase of screening when shifting from G-S ($\varepsilon=1$) to
G-BN ($\varepsilon_\mathrm{hBN}\approx{5}$), and to G-Gr, where
the Coulomb interaction is screened by the conducting substrate,
which can be viewed as a large effective~$\varepsilon$. Notably,
Eq.~(\ref{vnFOPT=}) also predicts $v_2>v_1$ for the same values of
$\varepsilon$ and~$B$ (as $C_2>C_1$), which would be the opposite
if one simply substituted $E\to{E}_n^0$ in Eq.~(\ref{vFOPT=}).
This is due to vertex corrections. According to
Eq.~(\ref{vnFOPT=}), $\delta{v}_{21}\equiv v_2-v_1 = (\alpha
c/4\varepsilon)(C_2-C_1) $, which also agrees with the trend seen
in Fig.~\ref{fig2}, the decrease of $\delta{v}_{21}^{exp}$ with
increasing~$\varepsilon$.

However, Eqs.~(\ref{vFOPT=}), (\ref{vnFOPT=}) fail to reproduce
quantitatively the data shown in Fig.~\ref{fig2}. This is with
regard to both the apparent amplitude of the slopes of
$v_{n}^{exp} \propto \ln(B)$ dependences as well as the observed
values of the relative shift $\delta{v}_{21}^{exp}$ between the
velocities associated with $L_1$ and $L_2$ Landau levels.

If one uses Eq.~(\ref{vnFOPT=}) with some adjustable
$\varepsilon_*$ (effective dielectric constant) instead of
$\varepsilon$, the slopes in Fig.~\ref{fig2} would correspond
respectively to $\varepsilon_*=3.9 \pm0.3$, $7.0 \pm0.5$ and
$12.0\pm1.0$ for G-S, G-BN and G-Gr, quite different from the
known $\varepsilon=1$ and $\varepsilon=5$ for G-S and G-BN. This
is not very surprising as the perturbative Eq.~(\ref{vnFOPT=})
does not have to work when the expansion parameter
$\alpha_\varepsilon$ (exceeding 2 for $\varepsilon=1$) is not
small. Fortunately, graphene offers another expansion parameter
which can control the perturbation theory even when
$\alpha_\varepsilon\sim{1}$. This parameter is identified as
$1/N$, where $N$ is the number of electronic species: $N=4$ for
graphene (the combined spin and valley degeneracy). In the $1/N$
expansion, an infinite number of terms of the perturbation theory
is re-summed to all orders in $\alpha_\varepsilon$, selecting only
those corresponding to the leading order in
$1/N$~\cite{Gonzalez1999}. The resulting series is equivalent to
RPA, and it was explicitly shown that the subleading contribution
is indeed small~\cite{Hofmann2014}. The $1/N$ expansion has also
been successfully used to describe the Coulomb renormalization of
the electron-phonon coupling constants~\cite{Basko2008}.

For the velocity renormalization at $B=0$, the $1/N$ expansion
boils down to the modification (depending on $\alpha$) of the
coefficient in front of the logarithm in
Eq.~(\ref{vFOPT=})~\cite{Gonzalez1999}. For moderate values of
$\alpha < 2.5$, typical for graphene, this modified coefficient
can be well approximated (with 1\% precision)
by~\footnote{Following the $1/N$ expansion~\cite{Gonzalez1999},
the exact expression  for the replacement of the  coefficient in
front of the logarithm in Eq.~(\ref{vFOPT=}) is: \newline
$\frac{\alpha_\varepsilon}4\to\frac{2}{\pi^2}
\left[1-\frac{1}{\alpha_\varepsilon}+\frac{2}{\pi\alpha_\varepsilon}\,
\frac{\arccos(\pi\alpha_\varepsilon/2)}
{\sqrt{1-(\pi\alpha_\varepsilon/2)^2}}\right]$}:

\begin{equation}\label{1overN=}
\frac{v_{0} \alpha_{\varepsilon}}{4} \to \frac{v_{0}
\alpha_{\varepsilon}}{4(1+1.28\alpha_\varepsilon)} =\frac{\alpha c
}{4 (\varepsilon +1.28\alpha c/ v_{0})}
\end{equation}
The above result can be seen as the added screening capacity,
$\varepsilon \to \varepsilon_{1/N} = \varepsilon +1.28 \alpha c/
v_{0}$,  by the graphene Dirac electrons themselves. Assuming
$v_0=0.88\times 10^6$~m/s (see below), we obtain
$\varepsilon_{1/N}=4.16$ for $\varepsilon=1$ (G-S) and 8.16 for
$\varepsilon=5$ (G-BN), which are quite close to the measured
values of~$\varepsilon_*$. Obviously, it is hard to reason in
terms of dielectric screening in case of G-Gr: a large
$\varepsilon_*=12$ ($\varepsilon_{1/N}\simeq\varepsilon$) found
for this graphene species must effectively account for efficient
screening by the conducting graphite substrate.

At this point we apprehend the slopes of the lines in
Fig.~\ref{fig2}. The apparent amplitude of the velocity shifts
$\delta{v}_{21}^{exp}$ remains to be analyzed. The measured values
are $\delta{v}_{21}^{exp}\simeq\{0.084,0.039\}\times{10}^6$~m/s
for G-S and G-BN, respectively, and we estimate that
$\delta{v}_{21}^{exp}\leq 0.01\times{10}^6 $ for G-Gr. On the
other hand, Eq.~(\ref{vnFOPT=}) gives $\delta v_{21}=(\alpha
c/4\varepsilon)(C_2-C_1)=\{0.110,0.022,0.009\}\times{10}^6$~m/s
for $\varepsilon=1$, $5$ and $12$, respectively, for G-S, G-BN and
G-Gr. The replacement $\varepsilon\to\varepsilon_{1/N}$ in
Eq.~(\ref{vnFOPT=}) results in an even worse agreement with the
experiment. Indeed, this replacement is valid only for the leading
logarithmic term, while the sub-logarithmic terms should be
calculated explicitly, and the simple combination
$C_n/\varepsilon$ will be replaced, generally speaking, by some
more complicated one. Such a calculation has not been performed,
to the best of our knowledge, and is beyond the scope of the
present paper.

In order to describe the whole set of experimental data, we assume
the following ansatz:
\begin{equation}\label{vnPHE=}
v_n= v_{0} + \frac{\alpha c}{4\varepsilon_{*}} ( \mathcal{L} - \ln
\frac{l_{B_0}}{l_B}) + \frac{\alpha c}{4\varepsilon_{\delta v}}
C_n
\end{equation}
where $\varepsilon_*=3.9,\,7,\,12$ at the leading logarithmic term
is in reasonable agreement with the $1/N$ expansion. In the
sub-logarithmic term we fixed $C_n$ to be the same as in
Eq.~(\ref{vnFOPT=}) and $\varepsilon_{\delta{v}}$ to depend only
on $\varepsilon$ (but not on~$n$). We do not have a proper
theoretical justification for this assumption but adopting it, and
setting $\varepsilon_{\delta{v}}=1.3,\,3.7,\,12$ in order to
reproduce the experimentally observed $\delta{v}_{21}$, we are
left with only two adjustable parameters, $v_0$ and $\mathcal{L}$.
Their best matching values are $v_0=0.88\times{10}^6\:\mbox{m/s}$
and $\mathcal{L}=4.9$, that is,
$W=(\hbar{v}_0/l_{B_0})^\mathcal{L}=3.1\:\mbox{eV}$, in fair
agreement with the bare velocity and the characteristic bandwidth
expected in graphene~\cite{Gillen2010}.

Concluding, using micro-magneto-Raman scattering spectroscopy, we
have studied inter Landau level excitations in graphene
structures, embedded in different dielectric environments.
Understanding the energies of inter LL excitations clearly falls
beyond the single particle approach (which refers to a simple
Dirac equation) but appears to be sound when the effect of
electron-electron interactions are taken into account. We confirm
that the electronic properties of graphene on insulating
substrates (weak dielectric screening) are strongly affected by
electron-electron interactions, whereas conducting substrates
favor the single particle behavior (graphene on graphite studied
here, but likely also graphene on metals~\cite{SooKim2009,
Coraux2008, Ruoff2009} and graphene on
SiC~\cite{Sadowski2006,Orlita2008}). The present experiment
together with the underlined theory show that the self-energy and
vertex (excitonic) corrections to zero-momentum inter LL
excitations do not cancel each other (breaking of the Kohn
theorem), as often speculated in the literature~\cite{Orlita2010}
and already invoked in one of the early magneto-spectroscopy
studies of graphene~\cite{Jiang2007}.

\begin{acknowledgments}
We thank Ivan Breslavetz for technical support and R. Bernard,  S.
Siegwald, and H. Majjad for help with sample preparation in the
StNano clean room facility, and P. Hawrylak for valuable
discussions. This work has been supported by the European Research
Council, EU Graphene Flagship, the Agence nationale de la
recherche (under grant QuanDoGra 12 JS10-001-01) and, the
LNCMI-CNRS, member of the European Magnetic Field Laboratory
(EMFL).

\end{acknowledgments}

%\bibliography{velocity}

%

\newpage
 ~
\newpage

%\begin{widetext}
\onecolumngrid

\begin{center}
{\large
\textbf{Supplementary Information for\\\vspace{2mm}
Landau level spectroscopy of electron-electron interactions in graphene}}\\ \vspace{3mm}
C. Faugeras,$^1$ S. Berciaud,$^2$ P. Leszczynski,$^1$ Y. Henni,$^1$ K. Nogajewski,$^1$ M. Orlita,$^1$ T. Taniguchi,$^3$\\
K. Watanabe,$^3$ C. Forsythe,$^4$ P. Kim,$^4$ R. Jalil,$^5$ A.K. Geim,$^5$ D.M. Basko,$^{6}$ and M. Potemski$^1$\\
\vspace{4mm}
{\small\textsl{$^1$Laboratoire National des Champs Magn\'{e}tiques Intenses, CNRS,\\
(UJF, UPS, INSA), BP 166, 38042 Grenoble Cedex 9, France\\
$^2$Institut de Physique et Chimie des Mat\'{e}riaux de Strasbourg and NIE, UMR 7504,\\
Universit\'{e}e de Strasbourg and CNRS, BP43, 67034 Strasbourg Cedex 2, France\\
$^3$National Institute for Material Science, 1-1 Namiki, Tsukuba, Japan\\
$^4$Department of Physics, Columbia University, New York, NY 10027, USA\\
$^5$School of Physics and Astronomy, University of Manchester, Manchester, M13 9PL, United Kingdom\\
$^6$Universit\'e Grenoble 1/CNRS, Laboratoire de Physique et de
Mod\'elisation des Milieux Condens\'es (UMR 5493), B.P. 166, 38042
Grenoble, Cedex 9, France\\}}
\end{center}

\twocolumngrid

Here we evaluate the first-order correction to the energy
$\hbar\omega_{-n,n}$ of the $L_{-1,1}$ inter-Landau-level
electronic excitation due to the unscreened Coulomb interaction
with the potential
\begin{equation}
U_{\vec{q}}=\frac{e^2}\varepsilon\,\frac{2\pi}{|\vec{q}|}\,
(1-\delta_{\vec{q},0}),
\end{equation}
where $e$ is the electron charge,
$\varepsilon$ is the background dielectric constant,
$\vec{q}$ is the two-dimensional wave vector,
and the last factor accounts for the neutralizing ionic
background. The first-order correction is given by the
expectation value of the interaction Hamiltonian in the
state
\begin{equation}
|L_{-n,n}\rangle=\sum_{p_x}
\sum_{v=K,K'}\sum_{\sigma=\uparrow,\downarrow}
\hat{c}^\dagger_{n,p_x,v,\sigma}
\hat{c}_{-n,p_x,v,\sigma}|0\rangle,
\end{equation}
where $\hat{c}^\dagger_{n,p_x,v,\sigma}$ is the creation
operator for an electron with the $x$~component of momentum
$p_x$ (we use the Landau gauge) on the $n$th Landau level
in the valley~$v$ and with the spin projection~$\sigma$.
$|0\rangle$ is the ground state of the system, corresponding
to all negative Landau levels filled, all positive Landau
levels empty, and the level $n=0$ half-filled. This expectation
value is given by
\begin{equation}
\hbar\omega_{-n,n}-\hbar\omega_{-n,n}^0=
\Sigma_n-\Sigma_{-n}+V_{-n,n},
\end{equation}
where $\Sigma_{\pm{n}}$ and $V_{-n,n}$ are the self-energy
and vertex corrections, respectively, given by
\begin{eqnarray}
&&\Sigma_{\pm{n}}=-\sum_{n'}{f}_{n'}
\int\frac{d^2\vec{q}}{(2\pi)^2}\,
\tilde{J}_{n',\pm{n}}(-\vec{q})\,\tilde{J}_{\pm{n},n'}(\vec{q})\,U_\vec{q},
\quad\\
&&V_{-n,n}=-\int\frac{d^2\vec{q}}{(2\pi)^2}\,
\tilde{J}_{n,n}(-\vec{q})\,\tilde{J}_{-n,-n}(\vec{q})\,U_\vec{q}.
\end{eqnarray}
Here $f_{n'}$ is the average occupation of the level $n'$
($f_{n'<0}=1$, $f_0=1/2$, and $f_{n'>0}=0$), and
\begin{eqnarray*}
&&\tilde{J}_{n\neq{0},n'\neq{0}}(\vec{q})=
\frac{J_{|n|,|n'|}(\vec{q})+\sign(nn')\,J_{|n|-1,|n'|-1}(\vec{q})}{2},\\
&&\tilde{J}_{n\neq{0},0}(\vec{q})
=\frac{J_{|n|,0}(\vec{q})}{\sqrt{2}},\quad
\tilde{J}_{0,n\neq{0}}(\vec{q})
=\frac{J_{0,|n|}(\vec{q})}{\sqrt{2}},\\
&&\tilde{J}_{0,0}(\vec{q})=J_{0,0}(\vec{q}),\\
&&J_{n,n'}(\vec{q})=(-1)^{n'+\min\{n,n'\}}
\sqrt{\frac{\min\{n,n'\}!}{\max\{n,n'\}!}}\,
e^{-q^2l_B^2/4}\\ &&\qquad\qquad
{}\times\left(\frac{q_x+iq_y}q\right)^{n-n'}
\left(\frac{ql_B}{\sqrt{2}}\right)^{|n-n'|}\\ &&\qquad\qquad
{}\times\mathrm{L}_{\min\{n,n'\}}^{(|n-n'|)}(q^2l_B^2/2),
\end{eqnarray*}
where
\[
\mathrm{L}_m^{(\alpha)}(\xi)=
\sum_{k=0}^m\frac{(-1)^k(m+\alpha)!}{k!\,(m-k)!\,(\alpha+k)!}\,\xi^k.
\]
is the associated Laguerre polynomial.

The vertex correction is straightforwardly evaluated,
\[
V_{-n,n}=-\frac{e^2}{4\varepsilon{l}_B}\int\limits_0^\infty
\frac{d\xi\,e^{-\xi}}{\sqrt{2\xi}}\,
\left[\mathrm{L}_n^{(0)}(\xi)+\mathrm{L}_{n-1}^{(0)}(\xi)\right]^2,
\]
which gives explicitly
\begin{equation}
\left\{V_{-1,1},\,V_{-2,2},\,V_{-3,3}\right\}=
-\frac{e^2}{\varepsilon{l}_B}\,\sqrt{\frac\pi{2}}
\left\{\frac{11}{16},\,\frac{145}{256},\,\frac{515}{1024}\right\}.
\end{equation}
For the self-energies of the lowest Landau levels we obtain
\begin{widetext}
\begin{eqnarray*}
&&\Sigma_0=-\frac{e^2}{\varepsilon{l}_B}\int\limits_0^\infty
\frac{d\xi\,e^{-\xi}}{\sqrt{2\xi}}
\left(f_0+\sum_{n=1}^\infty\frac{\xi^n}{2n!}\right)
=-\frac{e^2}{\varepsilon{l}_B}\sqrt{\frac\pi{2}}\left[f_0
+\frac{1}{2}\sum_{n=1}^\infty\frac{(2n)!}{4^n(n!)^2}\right],\\
%=-\frac{e^2}{l_B}\sqrt{\frac\pi{2}}\left[f_0-\frac{1}2
%+\frac{1}{2}\sum_{n=0}^\infty\frac{\Gamma(n+1/2)}{\sqrt\pi}\right],\\
&&\Sigma_{\pm{1}}=-\frac{e^2}{\varepsilon{l}_B}\sqrt{\frac\pi{2}}
\left[\frac{f_0}4+\sum_{n=1}^\infty\frac{2n-1/4\mp\sqrt{n}}{4(n-1/2)}\,
\frac{(2n)!}{4^n(n!)^2}\right],\\
&&\Sigma_{\pm{2}}=-\frac{e^2}{\varepsilon{l}_B}\sqrt{\frac\pi{2}}
\left[\frac{3f_0}{16}+\frac{15\mp{4}\sqrt{2}}{64}\right.+\\
&&\hspace*{1cm}+\left.\sum_{n=2}^\infty
\frac{64n^2-88n+9\pm\sqrt{2n}(36-32n)}{128(n-1/2)(n-3/2)}\,
\frac{(2n)!}{4^n(n!)^2}\right],\\
&&\Sigma_{\pm{3}}=-\frac{e^2}{\varepsilon{l}_B}\sqrt{\frac\pi{2}}
\left[\frac{5f_0}{32}+\frac{199\mp{4}\sqrt{3}(4+5\sqrt{2})}{512}\right.+\\
&&\hspace*{1cm}+\left.\sum_{n=3}^\infty
\frac{256n^3-928n^2+816n-75
\mp\sqrt{3n}(128n^2-416n+300)}{512(n-1/2)(n-3/2)(n-5/2)}\,
\frac{(2n)!}{4^n(n!)^2}\right].
\end{eqnarray*}
\end{widetext}
In each of the above expressions, the sum is divergent,
so the upper limit should be set to some large~$n_{max}$.
Then, each $\Sigma_{\pm{n}}\propto{n}_{max}$. The energy
differences diverge only logarithmically,
\begin{eqnarray}
&&\Sigma_{\pm{1}}-\Sigma_0=\pm\frac{e^2}{\varepsilon{l}_B}
\left(0.290240+\frac{1}{\sqrt{2}}\frac{\ln{n}_{max}}{4}\right),\\
&&\Sigma_{\pm{2}}-\Sigma_0=\pm\frac{e^2}{\varepsilon{l}_B}
\left(0.256384+\frac{\ln{n}_{max}}{4}\right),\\
&&\Sigma_{\pm{3}}-\Sigma_0=\pm\frac{e^2}{\varepsilon{l}_B}
\left(0.196799+\sqrt{\frac{3}2}\,\frac{\ln{n}_{max}}{4}\right),
\end{eqnarray}
where $f_0$ has been set to $1/2$.
Relating $n_{max}$ to the energy cutoff $W$ as
$\sqrt{2n_{max}}\,(\hbar{v}_0/l_B)=W$, we arrive at
Eq.~(2) of the main text with $C_1=-0.398$, $C_2=-0.197$,
$C_3=-0.193$.

%\end{widetext}
\end{document}